\documentclass[a4paper]{jpconf}
\usepackage{graphicx}

\begin{document}

\title{Quality Inspection Aspects of Hybrid Prototypes for the CMS Outer Tracker Upgrade at HL-LHC}

\author{A La Rosa$^1$, I Ahmed$^1$, J Almeida$^1$, G Blanchot$^1$, S Cooperstein$^2$,\\ I  Dominguez$^1$, A Honma$^3$, M Kovacs$^1$, A Zografos$^1$, \\for the CMS Tracker Group}

\address{$^1$ CERN, Geneva 23, CH-1211, Switzerland.} 
\address{$2$ University of California, San Diego, La Jolla, CA-92093, United States.}
\address{$3$ Brown University, 75 Waterman St, Providence, RI-02912, United States.}

\ead{alessandro.larosa@cern.ch}

\begin{abstract}
At the High Luminosity LHC (HL-LHC), the CMS experiment will need to operate at up to 200 interactions per 25\,ns beam crossing time and with up to 4000\,fb$^{-1}$ of integrated luminosity. To achieve the physics goals the experiment needs to improve the tracking resolution and the ability to selectively trigger on specific physics events.
The CMS tracker upgrade requires designing a new detector to cope with the HL-LHC conditions. The new Outer Tracker is based on two types of silicon modules (strip-strip and pixel-strip). Each module type has a few types of high-density interconnect hybrid circuits which house the front-end and auxiliary electronics. This paper introduces the technological choices for modules and hybrids and presents the quality inspection aspects of the first hybrid prototypes.
\end{abstract}


\section{Introduction}
\begin{figure}[h!]
\centering
\includegraphics[scale=0.34]{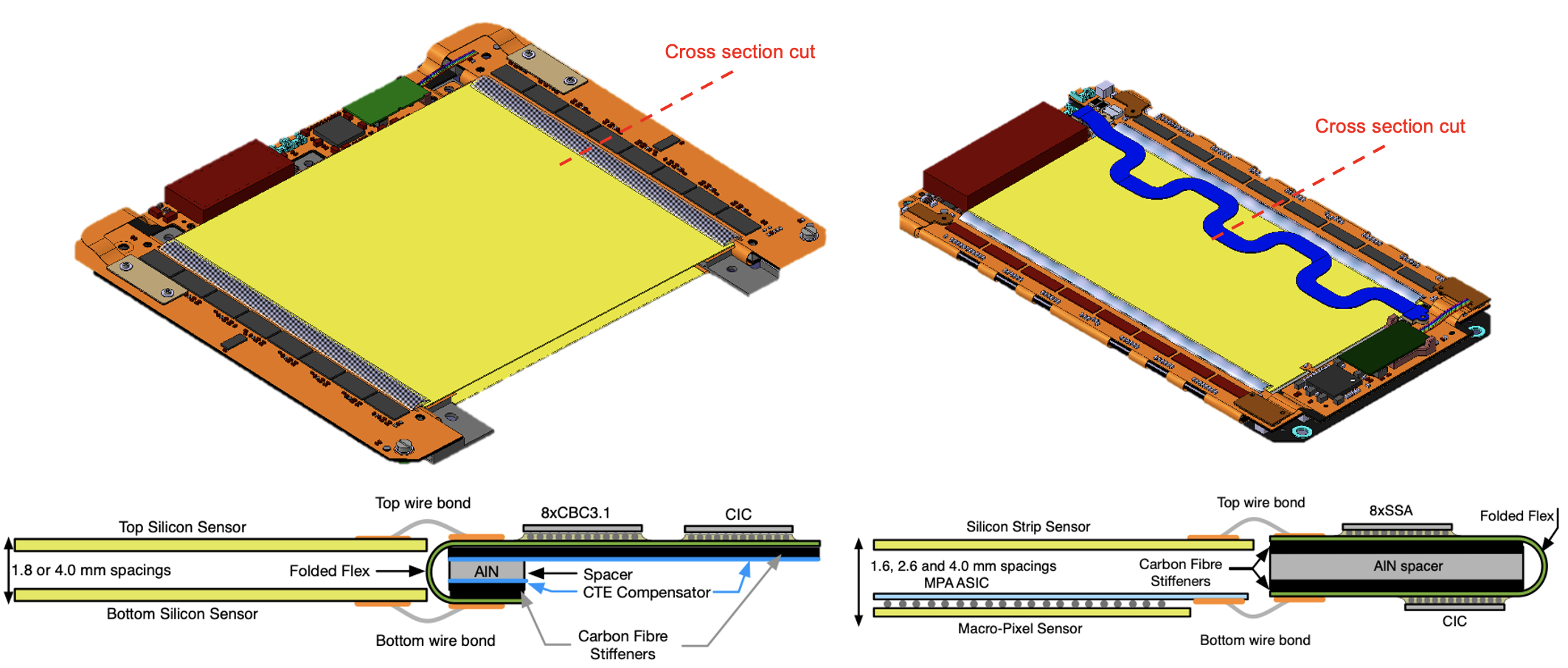}
\caption{The 2S module (left) and PS module (right) of the Outer Tracker. Shown are views of the assembled modules (top), and sketches of the front-end hybrid folded assembly and connectivity (bottom).}
\label{fig:modules}
\end{figure}

With the high-luminosity phase of the LHC (HL-LHC\,~\cite{HL-LHC}), the accelerator is expected to deliver an instantaneous peak luminosity of up to 7.5\,$\times$\,$10^{34}$\,cm$^{-2}$s$^{-1}$, and a total integrated luminosity up to 4000\,fb$^{-1}$  to the experiments. The CMS detector~\cite{CMS} will undergo a major upgrade for the HL-LHC, with an entirely new tracking detector consisting of an Inner Tracker and an Outer Tracker~\cite{TKTDR}.
The Outer Tracker (OT) will be based on two types of silicon modules  made of two sensors on top of each other, which are either two strip sensors (2S) or one macro-pixel sensor and one strip sensor (PS) as shown in Fig.\,\ref{fig:modules}. 
The OT module has the ability to autonomously select track segments (stubs) above a selected transverse momentum threshold and to send these to the back-end track finder system for pattern recognition and track fitting  before sending the final tracks to the Level-1 trigger~\cite{TKTDR}.


\section{Hybrid circuits}
\begin{figure}[h!]
\centering
\includegraphics[scale=0.36]{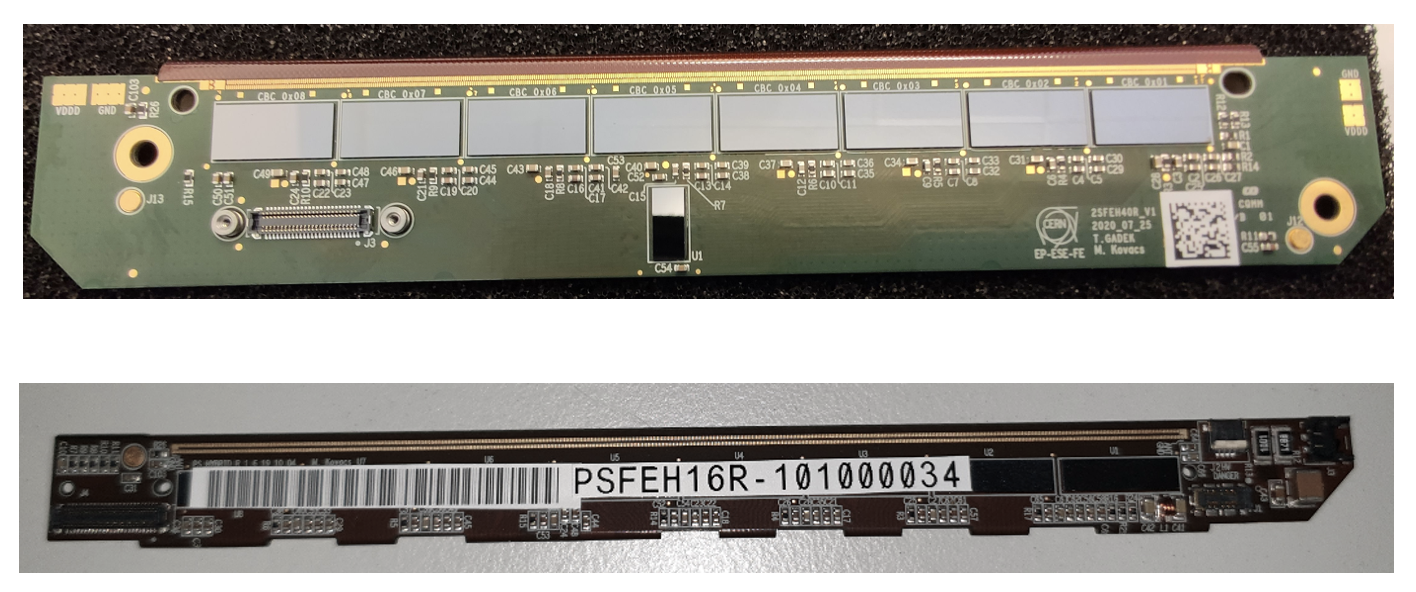}
\caption{Top view of the 2S front-end hybrid (top) and of the PS front-end hybrid (bottom).}
\label{fig:FEHs}
\end{figure}

As Fig.\,\ref{fig:modules} shows, each module type has several (three for the 2S module and four for the PS module) types of high-density interconnect hybrid circuits which house the front-end and auxiliary electronics.
The two sides of the sensors are wire-bonded to the front-end hybrids where the signals from the top and bottom sensor are routed to one readout chip to perform the track stub finding. This is possible by using a flexible hybrid which is folded over a Al-N spacer to allow for the same substrate height for bonding hybrid and sensor.
The front-end hybrids host eight readout chips (CMS Binary Chips, CBC~\cite{TKTDR} for the 2S module, and Short Strip ASIC, SSA~\cite{TKTDR} for the PS module) and one concentrator chip (CIC~\cite{TKTDR}, common for both module types). The readout chips are all bare flip-chip silicon dies and bump-bonded on the hybrids' surface leading to a very compact design, as shown in Fig.\,\ref{fig:FEHs}. The service hybrid hosts the LpGBT~\cite{TKTDR}, VTRx+ optical link~\cite{TKTDR}, DC-DC converters~\cite{TKTDR} and bias voltage distribution circuitry for the 2S module, while for space reasons, the PS module has two service hybrids, one for the optical system and one for the powering.
The hybrid circuits are implemented in 150\,$\mu$m thick four layer high density interconnect polyimide circuits featuring 42.5\,$\mu$m track width and spacing, 25\,$\mu$m copper filled microvias and a 250\,$\mu$m bump-bonding pitch. The digital data are transmitted on differential pairs matched to 90\,$\Omega$. In order to guarantee a rigid surface under the wire-bonding pads,  enable component assembly and allow heat transfer from the chips to the supporting structures for the cooling process, the hybrid circuits are laminated to a carbon fibre (CF) stiffener. Then, to obtain a flat hybrid without bowing effects,  coefficient of thermal expansion (CTE) compensator layers are introduced in the stack-up of the hybrid. This is made for the 2S front-end hybrids, the 2S service hybrids, and the PS readout hybrids that have asymmetric stack-up. The thickness  of the CTE compensators is optimised on the basis of the thickness and mechanical properties of the copper, polyimide, CF stiffener and adhesive materials used~\cite{Hybrids}. 


\section{Hybrid quality inspection}
\begin{figure}[h!]
\centering
\includegraphics[scale=0.3]{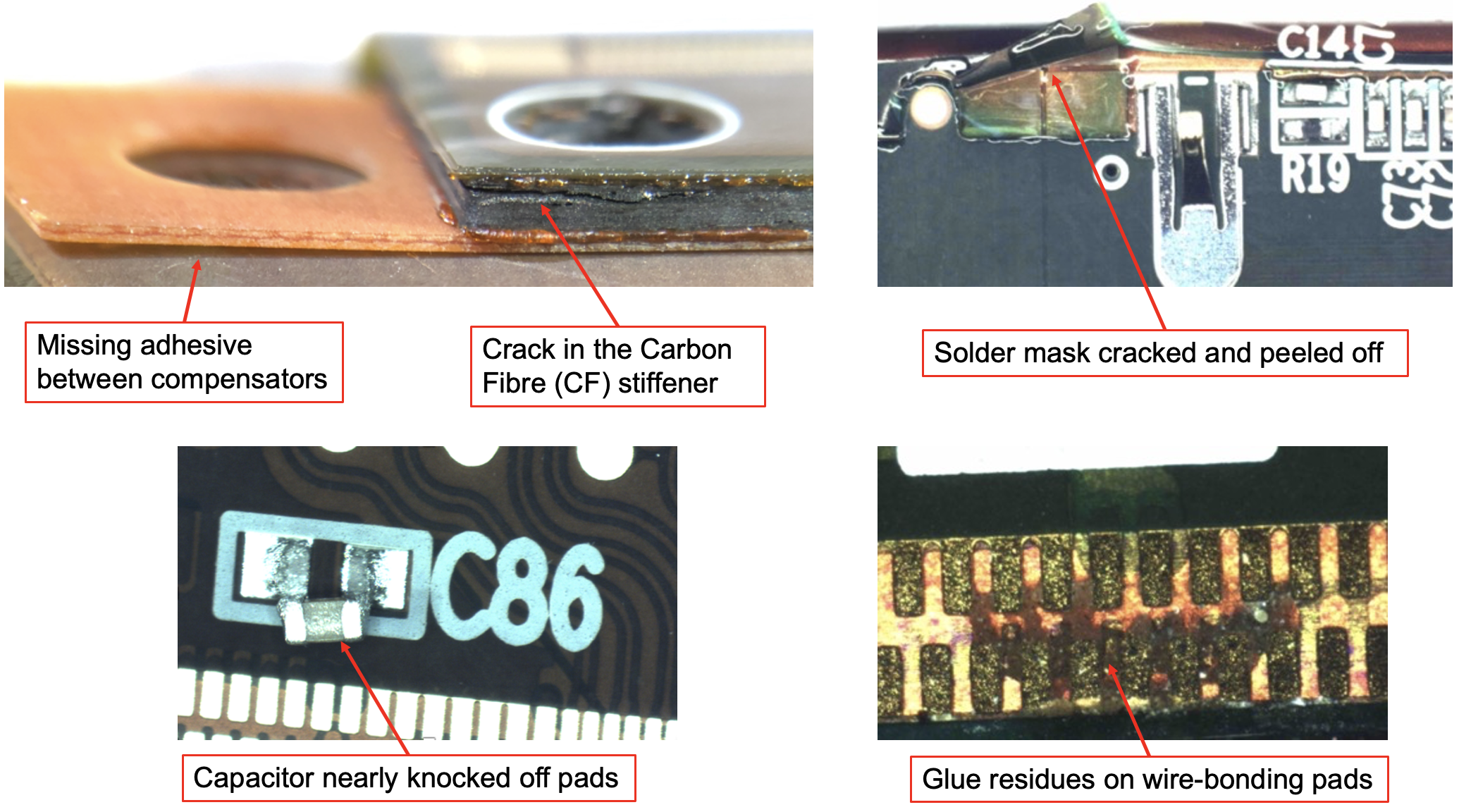}
\caption{Examples of problems spotted during the visual inspection of  some front-end hybrid prototypes.}
\label{fig:VI_1}
\end{figure}

\begin{figure}[h!]
\centering
\includegraphics[scale=0.55]{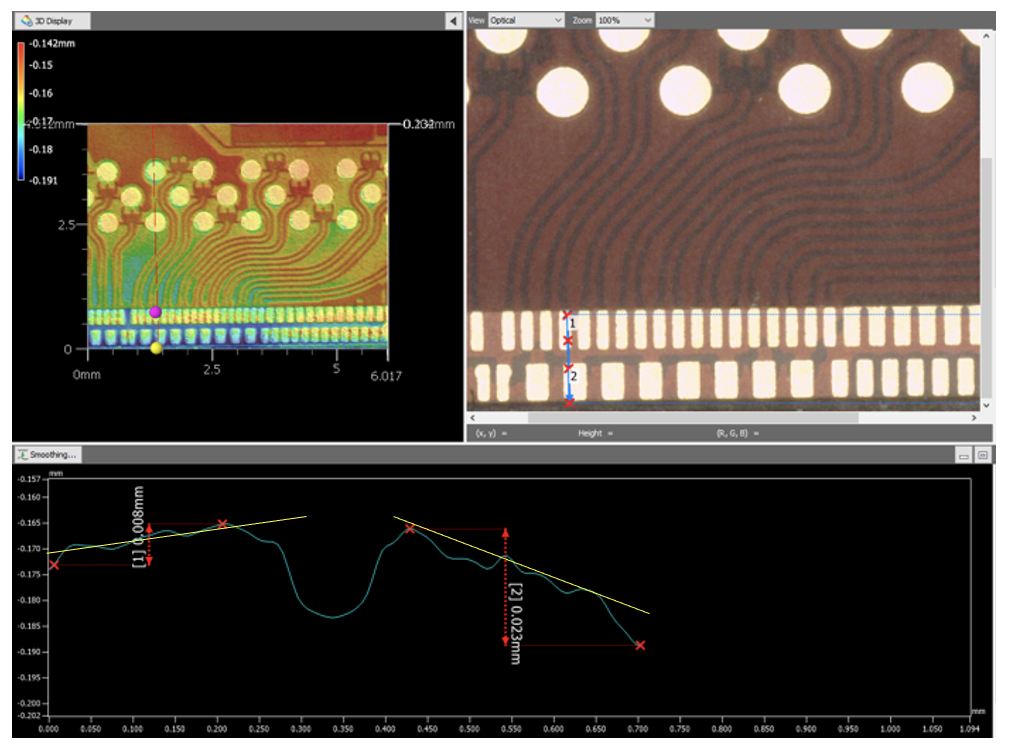}
\caption{Examples of local flatness measurements of bond pads. Shown are 3D (top left) and 2D (top right) views with the cross section cut for flatness measurement, and flatness profile (bottom).}
\label{fig:VI_2}
\end{figure}

The hybrid quality inspection process consists of a visual inspection and a  mechanical and electrical control. The visual inspection takes place using stereo microscopes with coaxial, ring and external lighting and consists of inspection and evaluation of the following aspects: the component soldering and placement quality; the alignment and adhesive quality of hybrid elements such as flex circuit, CF stiffener and Al-N spacers; the cleanliness of the circuit including the interior of the connectors; and where applicable the bond pad quality. \\
One of the main reasons for including the visual inspection of the hybrids in the quality control process is to ensure wire bondability of the hybrids to the sensors, as failure of wire-bonding can cause rejection of a full module. The bondability depends on many, mostly mechanical and chemical aspects, some of which can be detected visually and via  measurements.
The tolerances of the mechanical properties of the hybrids for module assembly are much stricter than in the past detector, and the reliability (mechanical and electrical) is critical given the operational conditions and long lifetime expected, thus the electrical and functional tests are not sufficient to ensure the quality of the hybrids. Examples of problems spotted during the visual inspection on some front-end hybrid prototypes are shown in Fig.\,\ref{fig:VI_1}.
In addition to the visual inspection, the bondability of the hybrids is evaluated by local and global flatness measurements of the bond pads and by wire-bonding and pull tests.
Figure\,\ref{fig:VI_2} shows an example where the bond pad flatness is measured along two neighbour pads. A non-flatness larger than 20\,-\,25\,$\mu$m could lead to a degradation of the bonding quality or, in  the worst case,  to the impossibility to bond on severely non-flat pads.
Also global flatness of the wire-bond pads has been verified on a sample of front-end hybrids. The global flatness is important for keeping the hybrid bond pad height uniform with respect to the sensor bond pads during the wire-bonding. In most of the cases, the variations observed were less than 100\,$\mu$m which should not cause bonding problems. These flatness variations should be monitored during production since variations larger than 100\,$\mu$m could lead to problems during module assembly. Last but not least, flex stretch (or shrinkage) of the front-end hybrids has been also measured by comparing the overall flex stretch (or shrinkage) to the layout dimensions. This effect could be critical for hybrid to sensor bond pad alignment and ASIC flip-chip solder bump alignment to the flex. In most of the cases, there was a good uniformity in the bond pad row length for different inspected hybrids. 
There were cases where there was a sizeable stretch in the flex of 100\,$\mu$m or more which could lead to wire-bonding problems such as skewed wire-bonding.

\noindent The acceptance of the hybrids requires to pass also electrical and functional tests that are carried out with a crate based on a multiplexing platform described in Ref.~\cite{SystemTest}. In addition, characterisation and stress testing of samples of the 2S front-end hybrid prototypes are performed at nominal operating temperature of $-$35$^{\circ}$C, including also thermal cycling between $-$35$^{\circ}$C and $+$20$^{\circ}$C to ensure reliable performance at the edge conditions of what can be expected during the CMS tracker operation and lifetime. 
Visual inspections are also performed before and after temperature cycling and no indication of mechanical damage from about 20 thermal cycles between $-$35$^{\circ}$C and $+$20$^{\circ}$C was observed.


\section{Summary}
The CMS Phase-2 tracker is an ambitious project that has to cope with the HL-LHC conditions. The new Outer Tracker will consist of about 13200 modules with about 29000 front-end hybrids and 18000 service, power and readout hybrids.
Thanks to the visual inspection, electrical control and characterisation made on the hybrid prototypes, valuable inputs were provided to the contractors to finalise the designs of the hybrids, assessing aspects related to mechanical assembly and interconnection properties. Further, visual inspection will be a crucial part of the quality control of the hybrids during production. The hybrid development project is overall on track
and pre-production is planned for the first half of 2022.\\

\end{document}